# Making Software FAIR: A machine-assisted workflow for the research software lifecycle

*Presentation submitted to the 19th International Conference on Open Repositories, June 3-6th 2024, Göteborg, Sweden (OR2024)*

*Petr Knoth, CORE, Knowledge Media institute, The Open University, petr.knoth@open.ac.uk*

*Laurent Romary, Inria, laurent.romary@inria.fr*

*Patrice Lopez, Science Miner, patrice.lopez@science-miner.com*

*Roberto Di Cosmo, Inria, rdicosmo@gmail.com*

*Pavel Smrz, Brno University of Technology, smrz@vut.cz*

*Tomasz Umerle, Polish Academy of Sciences, tomasz.umerle@ibl.waw.pl*

*Melissa Harrison, European Bioinformatics institute, mharrison@ebi.ac.uk*

*Alain Monteil, Inria, alain.monteil@inria.fr*

*Matteo Cancellieri, Knowledge Media institute, The Open University, matteo.cancellieri@open.ac.uk*

*David Pride, CORE, Knowledge Media institute, The Open University, david.pride@open.ac.uk*

## Abstract

A key issue hindering discoverability, attribution and reusability of open research software is that its existence often remains hidden within the manuscript of research papers. For these resources to become first-class bibliographic records, they first need to be identified and subsequently registered with persistent identifiers (PIDs) to be made FAIR (Findable, Accessible, Interoperable and Reusable). To this day, much open research software fails to meet FAIR principles and software resources are mostly not explicitly linked from the manuscripts that introduced them or used them.

SoFAIR is a 2-year international project (2024-2025) which proposes a solution to the above problem realised over the content available through the global network of open repositories. SoFAIR will extend the capabilities of widely used open scholarly infrastructures (CORE, Software Heritage, HAL) and tools (GROBID) operated by the consortium partners, delivering and deploying an effective solution for the management of the research software lifecycle, including: 1) ML-assisted identification of research software assets from within the manuscripts of scholarly papers, 2) validation of the identified assets by authors, 3) registration of software assets with PIDs and their archival.

## Audience

*librarians, repository managers, developers, researchers*

## Proposal

SoFAIR is a new 2-year international research project funded by the CHIST-ERA Open & Re-usable Research Data & Software Call. SoFAIR will extend the capabilities of existing widely used open scholarly infrastructures (CORE, Software Heritage, HAL) and tools (GROBID) operated by the consortium partners, delivering and deploying an effective solution for the management of the research software lifecycle. The consortium partners include: The Open University (operating CORE), INRIA (operating the HAL repository, Software Heritage and GROBID), the European Bioinformatics Institute (operating Europe PMC), Brno University of Technology and the Polish Academy of Sciences.

The ambition of SoFAIR is to:

1. Develop a **machine learning assisted workflow for software assets lifecycle** covering all the steps from 1) **identification of software mentions** in research manuscripts, 2) their **validation** by authors, 3) their **registration with PIDs** and **archival** if needed.
2. **Embed this workflow** into established scholarly infrastructures, making the solution available to the **global network of open repositories**, covering tens of millions of open access research papers originating from across more than 10k repository systems.

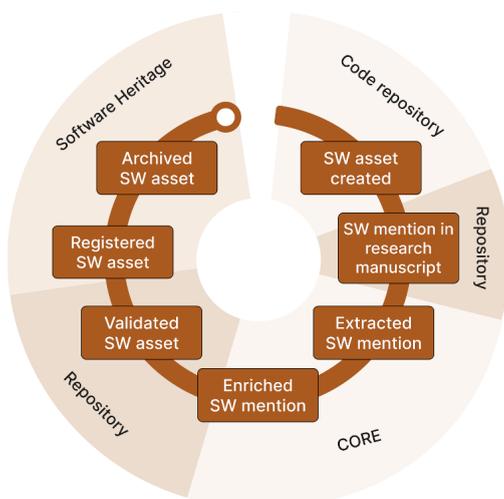

Figure 1: Software Asset Lifecycle

In the context of the SoFAIR project, the software asset lifecycle (Figure 1) begins with the creation of the research software asset itself. At this point this asset might be available in a code repository, such as GitHub. Once created, the first descriptions and mentions of, and links to, these research software assets can most often be identified in the full text of the associated research study. Building on existing and going beyond the state-of-the-art for software mentions extraction, we will develop the functionality to extract software mentions from research papers. These mentions then need to be enriched with additional descriptive metadata, e.g. by processing documentation from code repositories. Following extraction and enrichment, the next phase is to disambiguate and validate the discovered mentions. Our project addresses the validation by emplying the manuscript's authors. Once validated, an asset can be registered, assigned a persistent identifier (PID) and archived. This then enables repository systems storing research manuscripts to expose explicit links to software as part of their metadata.

We will build on established protocols, such as OpenAIRE Guidelines v4.0, RIOXX v3 and CodeMeta, to encode information about software assets and their links to research manuscripts establishing an interoperable and extensible workflow connecting open repositories (represented by the HAL repository), aggregators (represented by CORE) and software archives (represented by Software Heritage). The final solution will be applicable out-of-the-box to content from any open repository (supporting OAI-PMH) and will only require the installation of a repository-specific module to handle validation requests.

Specifically, SoFAIR project will leverage the following tools and infrastructures to deliver its objectives;

**GROBID:** is an open source machine learning library for extracting, parsing and re-structuring raw documents such as PDF into structured XML/TEI encoded documents with a particular focus on technical and scientific publications. Grobid has been developed since 2008, scales well to millions of documents and is widely used by scholarly infrastructures. recently, it has also been used for Open Science monitoring (Bssinet et al., 2023). Built on top of GROBID, the Softcite module is performing state-of-the-art software mention recognition (Lopez et al., 2021). **SoFAIR will extend Softcite with models for multidisciplinary identification of software mentions from research manuscripts, including their disambiguation and enrichment.** We expect to achieve a greater than 80% performance (f1-measure) in a multidisciplinary context. We will then describe these identified software assets using CodeMeta guidelines, following the recent EOSC recommendation [EOSC SIRS report, European Commission, 2020].

**CORE (core.ac.uk):** CORE (Knoth et. al, 2023, Knoth & Zdrahal, 2012) is a comprehensive bibliographic database of the world's scholarly literature and a repository aggregator used by around 30 million monthly active users. It is a not-for-profit service dedicated to the open access mission and subscriber to POSI principles [Bilder,2015]. **CORE will enable the application of the developed ML-assisted workflow on both pre-existing and new open access content from any open repository in the world,** paving the way for FAIRification of large numbers of software assets from over 30 million open access papers hosted by CORE. Specifically, CORE will 1) deploy the extended GROBID software and run it as part of its ingestion pipeline, 2) route newly identified but not yet validated software assets (via the free-to-use CORE Repository Dashboard service) to institutional repository managers who will be, in turn, able to approve their automated routing for validation by authors located at their institution. CORE will also 3) adapt its ingestion mechanism to support the explicit linking of research outputs with software, building on OpenAIRE, Rioxx and Signposting protocols, delivering an interoperable and machine-actionable solution for linking papers with software.

**HAL** is the primary open repository in France for scientific articles, hosting over a million full text research papers. It also serves as a repository for Inria researchers, in the context of its deposit mandate. **HAL will be used as a best practice example of a repository participating in the SoFAIR workflow**, it will 1) participate in the routing of identified software mentions coming from CORE for validation by Inria authors and 2) adapt its repository software to expose (over OAI-PMH) links between research manuscripts and software assets used in their creation.

**Software Heritage:** collects, preserves, and shares all software that is publicly available in source code form. The Software Heritage archive is the largest public collection of source code in existence and currently hosts more than 13 billion individual files from 192 million different projects. **Software Heritage will be embedded into the introduced workflow to support the registration of newly identified and validated software assets with PIDs and their subsequent archival.** All the communication between the services and tools mentioned above will be realised through the use of relevant open protocols, including Codemeta, Rioxx, OpenAIRE Guidelines, etc.

## 1.3 The SoFAIR approach

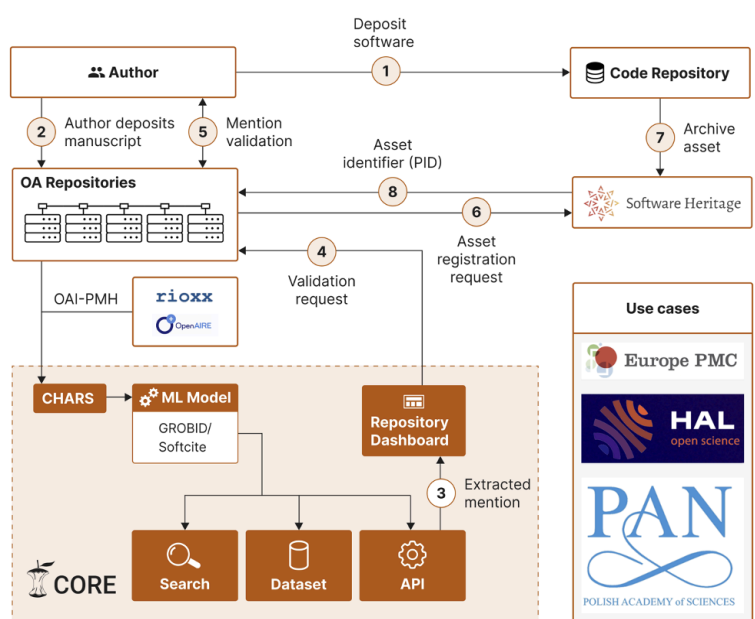

Figure 2: SoFAIR Workflow Diagram

The SoFAIR workflow shows how stakeholders, tools and infrastructures work in combination. An author deposits a piece of research software in a code repository [1]. The author then deposits a manuscript that contains either explicit or implicit mentions of that software [2]. The research paper is then harvested from the repository by CORE and software mentions extracted from the full text research paper using extended state-of-the-art ML tools (GROBID / Softcite (Lopez, et al., 2021)) [3]. Via the CORE Repository Dashboard, a request to validate the extracted mentions is made available to the repository [4] and, with the authorisation of the repository manager, routed to the author (e.g. by means of an email notification) who validates this request [5]. Once validated, the repository issues an asset registration request to Software Heritage [6] who permanently archive the new software asset [7] and issue a permanent identifier for the new asset and send this back to the repository [8].

Following this, repositories will be able to expose information linking software assets with research outputs that mention them within their OAI-PMH feed to aggregators and do so in an interoperable fashion. The efficacy of the solution will be validated within two disciplinary use cases: 1) a life sciences use case conducted in cooperation with Europe PMC and a 2) digital humanities use case conducted in cooperation with DARIAH. An additional multidisciplinary use case will be conducted in cooperation with the HAL repository. The key innovations of the project are:

**A novel machine-assisted workflow for software assets lifecycle management**

The most popular code repositories used by researchers, including GitHub and GitLab, are currently largely disconnected from the scholarly publishing workflow. As these tools are predominantly privately owned and the academic community exercises little control and influence over their design. Despite recent efforts, such as (Katz, 2021) introducing *citation files* integrated in GitHub, current practices for referencing software in research papers are typically limited to inserting a URL and/or a name of the software used in the research manuscript. Furthermore, Howison & Bullard (2016) found that 63% of mentions to software are informal (mentions without references), the remaining 37% being the software name associated with a paper reference and never to a software reference or to a software PID. SoFAIR offers a significant innovation to this practice by means of providing a **machine-assisted solution addressing the whole software assets management lifecycle**, **from the point of the identification of the software mention from within the research manuscript all the way to the registration and archival of the software asset.**

**New machine learning models for software mentions extraction and disambiguation**

Due to the software citation advocacy in the last decade, automatic recognition of software mentions has attracted considerable interest. (Kruger et al., 2020) presents first approaches, based mainly on gazetteers and rules. While ML approaches promised higher accuracy, they were first limited by the lack of manually annotated training data - the largest public dataset until 2020 was limited to 85 annotated documents. With the development of large annotated gold corpus, Softcite (Du et al, 2021), 5000 articles, and SoMeSci (Schindler et al. 2021), 1367 articles, deep learning approaches have become recently possible.

(Lopez et al. 2021) is the only ML recognizer adapted to the actual distribution of software mentions, which are extremely sparse, relying intensively on negative sampling techniques, while the others were trained only with positive examples. These recognizers are, however, limited by the poor multidisciplinary coverage of the training data. (Lopez et al. 2021) shows that the recognition performance falls by around 20 points F-score on a new scientific domain. In addition, after the extraction of millions of mentions, it is very challenging to deduplicate and aggregate reliably mentions to the same software, because of 1) the ambiguity and the incompleteness of extracted mentions and 2) the lack of central reference metadata information for software, which are fragmented in various places (code repositories, software and data registries, package managers, etc.). **SoFAIR will focus precisely on these two main issues by extending the training data and the Softcite models to new domains and by experimenting with recent supervised machine learning techniques for entity disambiguation, in particular using graph-based similarity techniques for entity matching/alignment.**

**Scalable application of the technology across open repositories and relevance to both pre-existing and new software assets**

The integration and deployment of the machine learning extraction software within CORE delivers a repository-centric solution to the problem of extraction and validation of software mentions from open access papers. Our commitment to the use of open protocols for communication between all participating services will provide an interoperable solution with the widest possible applicability across the open scholarly ecosystem, while at the same time providing further opportunities for extensibility and flexibility. This is in contrast to most previous approaches which tend to be limited to 1) the provision of source code only and 2) applicability of the approach only in a single domain. Another key characteristic of the SoFAIR approach is that **it can be applied consistently to both pre-existing and newly deposited content from anywhere within the global open repositories network**. **SoFAIR integrates key functionality within open scholarly infrastructures (CORE, Software Heritage, repositories) providing its service where users already are instead of asking them to come to a new platform.**

## Conclusions

SoFAIR will improve and semi-automate the process for identifying, describing, registering and archiving research software, ensuring it has received a Software Heritage persistent identifier (SWH-ID). The solution extensively builds on enhancing existing widely used open source tools (GROBID) and open scholarly infrastructures (CORE, Europe PMC, HAL, Software Heritage), which are operated by the consortium members. This ensures fast and wide adoption of the project's outputs across the global scholarly ecosystem of open repositories, offering tangible pathways to impact. The project closely aligns with the FAIR agenda, Horizon Europe initiatives on open research data and software, and European infrastructures including EOSC, CLARIN and DARIAH. It will apply a number of open protocols and recommended scholarly metadata standards such as, but not limited to, CodeMeta, OpenAire Guidelines, Rioxx and Signposting, to describe identified, enriched and curated research software assets as interoperable metadata and connect them to research manuscripts through explicit links. By lowering the barriers for research software to be made FAIR, this project will drive a step-change in the way research software can be found, accessed and referenced. This will facilitate not only software reuse and reproducibility but will also contribute to incentivising the creation of research software.

*— End of Page 3 of the proposal —*

# References (if applicable)